\documentclass{article}

\PassOptionsToPackage{numbers,compress}{natbib}

\usepackage[preprint]{neurips_2026}

\usepackage[utf8]{inputenc} %
\usepackage[T1]{fontenc}    %
\usepackage{url}            %
\usepackage{booktabs}       %
\usepackage{amsfonts}       %
\usepackage{nicefrac}       %
\usepackage{microtype}      %

\usepackage{subcaption} 
\usepackage{etoc}
\usepackage{enumitem}
\usepackage{graphicx}
\usepackage{amsmath}
\usepackage[normalem]{ulem}
\usepackage{multirow}
\usepackage{caption}
\usepackage{wrapfig}
\usepackage{stmaryrd}
\usepackage[table]{xcolor}
\usepackage[most]{tcolorbox}
\usepackage[colorlinks, linkcolor=mydarkred, citecolor=myblue]{hyperref}
\definecolor{mydarkblue}{rgb}{0,0.08,0.45}
\definecolor{mydarkred}{rgb}{0.6,0,0}
\definecolor{myblue}{HTML}{268BD2}
\definecolor{mygreen}{HTML}{658354}
\definecolor{results}{RGB}{220, 230, 240}

\newcommand{\name}{UniRank}

\title{UniRank: Unified List-wise Reranking via Confidence-Ordered Denoising}

\author{
    Pengyue Jia$^{1}$, HailanYang$^{2}$, Shuchang Liu$^{2}$, Xiaobei Wang$^{2}$, Wanyu Wang$^{1}$,\\ \textbf{Xiang Li}$^{2}$, \textbf{Yongqi Liu}$^{2}$, \textbf{Kaiqiao Zhan}$^{2}$, \textbf{Kun Gai}$^{2}$, \textbf{Xiangyu Zhao}$^{1}$ \\
    $^1$City University of Hong Kong, $^2$Kuaishou Technology \\
    \texttt{jia.pengyue@my.cityu.edu.hk,xianzhao@cityu.edu.hk }
}

\begin{document}

\maketitle

\begin{abstract}

List-wise reranking arranges a request-specific pool of candidate items into an ordered slate that maximizes user satisfaction.
Existing generative rerankers fall into two paradigms:
Autoregressive (AR) rerankers construct the slate left to right and capture inter-item dependencies in the exposure list, but they suffer from
error propagation because early mistakes affect subsequent slots.
Non-autoregressive (NAR) rerankers predict all slots in parallel and avoid error propagation, but they weaken inter-item interaction modeling
under a slot independence assumption.
This raises a central question: is there a unified architecture that combines the strengths of both paradigms and delivers stronger reranking
performance?
We answer this question with \textbf{\name}, \textit{a unified list-wise reranking framework whose inference time variants recover AR and NAR
rerankers as special cases}.
\name~integrates bidirectional slate modeling into an iterative denoising process and fills the most confident slot at each step.
To instantiate this framework for reranking, we introduce the Task Grounded Diffusion Interface (TGD), which performs denoising at the item level and restricts prediction to the request-specific candidate pool.
TGD aggregates each item's semantic tokens into a single item embedding and scores each slot directly against the candidate pool.
Experiments on Amazon Books, MovieLens-1M, and an industrial short video dataset show that \name~consistently outperforms state-of-the-art baselines.
Online A/B tests on a real-world industrial platform further validate its effectiveness, yielding significant improvements of +0.159\% in user average app-time and +1.016\% in share-rate.
\end{abstract}

\section{Introduction}
\label{sec:intro}

Recommender systems on modern online content-sharing platforms typically follow a multi-stage cascade~\cite{covington2016deep,liu2025recflow} of retrieval, ranking, and reranking.
As the final stage of this cascade, reranking~\cite{ai2018learning,pei2019personalized,liu2022neural} takes a request-specific local candidate set produced by earlier stages and arranges the items into an ordered slate for user exposure.
Unlike retrieval and ranking, which often evaluate items independently, reranking must model \emph{inter-item interactions}, including substitution, complementarity, and diversity.
For example, placing two near-identical items next to each other can reduce slate diversity, while placing complementary items together may improve user satisfaction.
The central challenge in reranking is therefore to generate an ordered slate that captures these interactions while respecting the candidate pool returned by earlier stages.

\begin{figure}[t]
    \centering
    \includegraphics[width=\linewidth]{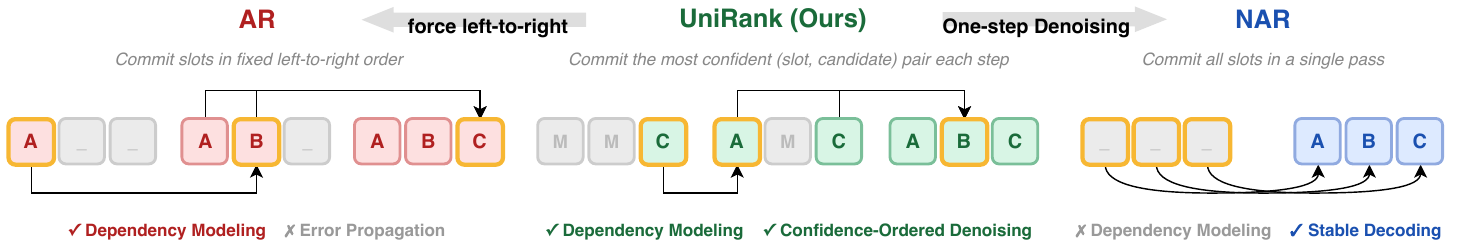}
    \caption{Comparison of AR, NAR, and \name~reranking paradigms. AR reranking (left) decodes the slate in a fixed order, while NAR reranking (right) predicts all positions in parallel. \name~(middle) uses confidence-ordered iterative denoising with bidirectional slate context; forcing the denoising order to follow slate positions recovers AR decoding, while using a single parallel denoising step recovers NAR decoding.}
    \label{fig: rerank_paradigm}
\end{figure}

Recent work frames reranking as a sequence generation problem $P(S|u,C)$, where the model constructs an exposure slate $S$ from the user context $u$ and a request-specific candidate pool $C$.
The majority follow two paradigms:
\emph{Autoregressive (AR) rerankers}~\cite{bello2018seq2slate,liu2023gfn4list,zhang2026goalrank} factorize the slate in a fixed left-to-right order and condition each item based on previously generated items.
This factorization captures dependencies among exposed items, but the causal attention mask prevents each slot from using information from later slots, and an early mistake can propagate irreversibly to subsequent slots~\cite{yang2026bear}.
\emph{Non autoregressive (NAR) rerankers}~\cite{pei2019personalized,ren2024non,mao2026denoising} predict all slots in parallel and avoid this error propagation, but they usually rely on a slot-wise independence assumption that weakens dependency modeling within the exposure slate.
A recent diffusion-based reranker~\cite{lin2024discrete} introduces iterative slate refinement, but it operates after an initial complete list has already been given.
Thus, it corresponds more closely to refinement over a starting slate than to adaptive slate construction from uncertain positions.
As illustrated in Figure~\ref{fig: rerank_paradigm}, existing generative rerankers therefore expose a tradeoff between \textbf{exposure dependency modeling} and \textbf{generation stability}.

This raises the central question of this paper: \emph{can we design a unified architecture that combines the strengths of both paradigms, modeling bidirectional inter-item interactions across a slate of exposure while mitigating the generation instability caused by fixed-order error propagation?}

We answer this question with \name, a unified list-wise reranking framework that constructs slates via confidence-ordered iterative denoising.
Starting from an all mask slate, \name~uses bidirectional slate modeling to score candidate assignments for each unfilled position, commits the most confident position-item assignment, and repeats this process until the slate is complete.
This design lets every prediction use the current slate context while avoiding a fixed left-to-right commitment order.
Under different inference schedules, the same framework recovers both paradigms: 
1) Forcing the denoising order to follow left-to-right gives an AR reranker, while 2) collapsing the process into all-slot parallel generation steps gives an NAR reranker.
Thus, \name~provides a unified inference view that connects AR decoding, NAR decoding, and confidence-ordered iterative decoding.

To instantiate \name~for reranking, the diffusion process must also match the item-level modeling and candidate-constrained nature of the task.
Mechanically, we introduce the Task Grounded Diffusion (TGD) interface, which contains two modules:
1) \emph{Semantic Fusion Layer (SFL)} aggregates each item's semantic representations into a single item embedding, so each denoising position corresponds to one item rather than a sequence of tokens.
2) \emph{Latent Pool Selection (LPS)} replaces vocabulary level decoding with dot product scoring against the local candidate pool, so each prediction is made directly over legal candidates for the current request. Extensive experiments on Amazon Books, MovieLens 1M, and an industrial short video dataset show relative improvements of up to 5.4\% in Precision, 4.4\% in NDCG, and 6.4\% in MAP over the strongest baseline. Online A/B tests on a real-world industrial platform further confirm real-world effectiveness, with +0.159\% in user average app-time and +1.016\% in share-rate.
In summary, our main contributions are as follows:
\begin{itemize}[leftmargin=*]
  \item We present a unified reranking architecture, \name, that combines bidirectional slate modeling with confidence-ordered iterative denoising, and recovers AR and NAR rerankers as inference-time special cases.
  \item We introduce the Task Grounded Diffusion (TGD) Interface, consisting of the Semantic Fusion Layer (SFL) and the Latent Pool Selection (LPS), which instantiates diffusion reranking at the item level and within the request-specific candidate pool.
  \item We conduct experiments on three datasets, and the results show that \name~outperforms state-of-the-art reranking baselines, with ablations validating the contribution of the proposed components.
\end{itemize}

\section{Methodology}\label{sec:method}

\begin{figure}
    \centering
    \includegraphics[width=\linewidth]{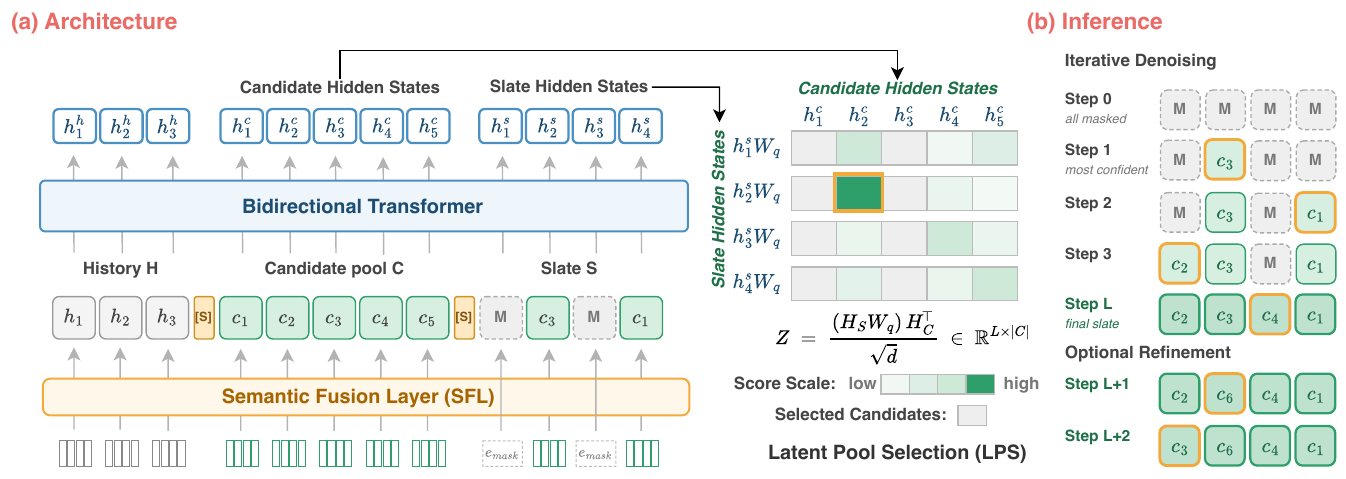}
    \caption{Overview}
    \label{fig:overview}
\end{figure}

\subsection{Problem Setup}
\label{sec:method:setup}

A reranking request consists of a user input $u$ and a candidate pool $\mathbf{C}=\{c_1, \dots, c_{|C|}\}$ provided by an upstream ranker/retriever.
The reranker solves a conditional policy $p_\theta\!\left(\mathbf{S} \mid u, \mathbf{C}\right)$ that produces an ordered slate $\mathbf{S}=[s_1, \dots, s_L]\in \mathbf{C}^L$ of $L$ distinct items.
Under sequential modeling, the user input $u$ is typically converted into a sequential representation $\mathbf{H}$, which may include but is not limited to user history and other profile features.
Training seeks parameters $\theta$ that maximize the expected list-wise utility of the generated slate:
\begin{equation}
\theta^{\star} \;=\; \arg\max_{\theta}\; \mathbb{E}_{(u, \mathbf{C})\sim \mathcal{D}}\;\mathbb{E}_{S \sim p_\theta(\cdot \mid u, \mathbf{C})}\big[\,U(\mathbf{S})\,\big],
\label{eq:rerank-objective}
\end{equation}
where $\mathcal{D}$ is the distribution of requests and $U$ is a slate level metric (e.g., NDCG@L or sum of clicks) that scores the full ordered slate $\mathbf{S}$. 

\subsection{\name: Unified Confidence-Ordered Denoising}
\label{sec:method:unirank}

In \name, we instantiate $p_\theta$ with a masked discrete diffusion model whose denoising backbone attends jointly to the history, the candidate pool, and a partially filled slate, modeling reranking as masked discrete diffusion over slate positions (Figure~\ref{fig:overview}).

Formally, let $\mathbf{x}_0 = (\mathbf{x}_0^{(1)}, \dots, \mathbf{x}_0^{(L)})$ denote the ground truth slate, where each $\mathbf{x}_0^{(j)}$ is the item representation assigned to slot $j$. 
Given a mask ratio $\alpha \in (0, 1]$, we draw a subset $\mathcal{M} \subseteq \{1, \dots, L\}$ of size $\lceil \alpha L \rceil$ uniformly at random and form the corrupted slate $x_\alpha$ by
\begin{equation}
\mathbf{x}_\alpha^{(j)} \;=\;
\begin{cases}
\mathbf{e}_{\text{mask}}, & j \in \mathcal{M}, \\
\mathbf{x}_0^{(j)}, & j \notin \mathcal{M},
\end{cases}
\label{eq:forward}
\end{equation}
where each slot is either replaced by the mask embedding or kept as its original item representation. Masking is applied only to the output slate $\mathbf{S}$, while the user history $\mathbf{H}$ and candidate pool $\mathbf{C}$ remain clean throughout. 

The reverse model $f_\theta$ is a shared $\mathbf{x}_0$ predictor parameterized by a stack of bidirectional Transformer blocks, implementing the distribution $p_\theta(\,\cdot \mid \mathbf{x}_\alpha, \mathbf{H}, \mathbf{C})$ that predicts the item generation probability (i.e., item confidence) at every masked slot (details in Section~\ref{sec:method:tgd}). We introduce a separator embedding $\mathbf{e}_{\text{sep}}$ to delimit the history, candidate, and slate in the input sequence $[\,\mathbf{H}\,\mid\,\mathbf{e}_{\text{sep}}\,\mid\,\mathbf{C}\,\mid\,\mathbf{e}_{\text{sep}}\,\mid\,\mathbf{x}_\alpha\,]$ under full self-attention. Since the random masking of Eq.\eqref{eq:forward} produces $\mathbf{x}_\alpha$ for all mask ratios $\alpha$, \name~can learn a single shared denoiser across all denoising steps.
This design enables full bidirectional attention over the slate at every prediction step, while deferring the commit order to inference time (e.g., committing the slot-item pair with the maximum confidence at each step).

\subsection{Task Grounded Diffusion Interface}
\label{sec:method:tgd}

To align with the item-level and candidate-constrained nature of the reranking task, we introduce the Task Grounded Diffusion (TGD) interface, comprising the Semantic Fusion Layer (SFL) and the Latent Pool Selection (LPS) module.

\subsubsection{Semantic Fusion Layer}
\label{sec:method:fusion}

Following recent generative recommendation work~\cite{rajput2023recommender,nie2025llada,hou2025rpg}, we represent each item by a semantic ID (SID) of $M$ discrete tokens $(\text{sid}_{i,1},\dots,\text{sid}_{i,M})$ drawn from $M$ parallel codebooks, sharing a token embedding table $\mathbf{E} \in \mathbb{R}^{V \times d}$ across all codebook positions. For each item $i$, SFL aggregates its $M$ SID token embeddings via concatenation and projection into a single item representation $\mathbf{e}_i$, so that each denoising position corresponds to exactly one item:
\begin{equation}
\mathbf{e}_i \;=\; \mathbf{W}_{\text{agg}} \,\big[\,\mathbf{E}[\text{sid}_{i,1}]\,\Vert\,\cdots\,\Vert\,
\mathbf{E}[\text{sid}_{i,M}]\,\big] + \mathbf{b}_{\text{agg}}, \qquad 
\mathbf{W}_{\text{agg}} \in \mathbb{R}^{d \times Md},\; \mathbf{b}_{\text{agg}} \in \mathbb{R}^{d}.
\label{eq:sid-fusion}
\end{equation}
At denoising step $k$ with mask ratio $\alpha_k$, the model input is $[\,\mathbf{H}\,\mid\,\mathbf{e}_{\text{sep}}\,\mid\,\mathbf{C}\,\mid\,\mathbf{e}_{\text{sep}}\,\mid\,\mathbf{x}_{\alpha_k}\,]$, where $\mathbf{x}_{\alpha_k}$ is a partially masked slate and $\mathbf{e}_{\text{mask}} \in \mathbb{R}^d$ is a learnable embedding assigned to each masked slot.

\subsubsection{Latent Pool Selection}
\label{sec:method:lps}

Latent Pool Selection replaces vocabulary level decoding with direct selection from the request-specific candidate pool.
Let $\mathbf{H}_S \in \mathbb{R}^{L \times d}$ collect the slate slot hidden states returned by the $f_\theta$, and $\mathbf{H}_C \in \mathbb{R}^{|\mathbf{C}| \times d}$ collect the candidate hidden states from the same forward pass. 
We compute the slot to candidate score matrix as
\begin{equation}
\mathbf{Z} \;=\; \frac{(\mathbf{H}_S \mathbf{W}_q)\, \mathbf{H}_C^{\!\top}}{\sqrt{d}} \;\in\; \mathbb{R}^{L \times |\mathbf{C}|},
\label{eq:latent-pool-score}
\end{equation}
where $\mathbf{W}_q \in \mathbb{R}^{d \times d}$ is a learnable query projection. 
The distribution for slot $j$'s item generation is $p_\theta(\,\cdot \mid \mathbf{x}_\alpha, \mathbf{H}, \mathbf{C}) = \operatorname{softmax}(\mathbf{Z}_{j,:})$, whose support is exactly the current candidate pool. This design makes candidate constraints a part of the model architecture rather than a post hoc correction process.

\subsection{Optimization}
\label{sec:method:opt}

We train the model with a denoising surrogate that predicts target items at masked slots, optimizing a label-reweighted cross-entropy over masked positions. At each training step, we draw a mask count $n \sim \operatorname{Uniform}\{1, \dots, L\}$ and mask $n$ slate slots uniformly at random following Eq.\eqref{eq:forward}; the remaining $L-n$ slots keep their ground-truth item embeddings. The induced mask ratio $\alpha = n/L$ is therefore uniformly distributed across training steps.
For each masked slot $j$, let $\ell_j \in \{1,\dots,|\mathbf{C}|\}$ be the index of the ground-truth item $\mathbf{x}_0^{(j)}$ in the candidate pool $\mathbf{C}$. The per-example loss is
\begin{equation}
\mathcal{L} \;=\; \frac{1}{L}\sum_{j \in \mathcal{M}} 
\frac{r_j}{\alpha}\,\operatorname{CE}\!\Big(
\operatorname{softmax}(\mathbf{Z}_{j,:}),\;\ell_j\Big),
\label{eq:loss}
\end{equation}
where $\mathcal{M}$ is the set of masked slots and $\mathbf{Z}_{j,:}$ is the corresponding row of Eq.\eqref{eq:latent-pool-score}. The factor $1/\alpha$ keeps the expected per-slot gradient contribution invariant to the sampled mask ratio.
The factor $r_j$ provides label-aware reweighting that allows the same framework to incorporate richer feedback signals without changing the denoising architecture:
\begin{itemize}[leftmargin=*]
  \item When $r_j \equiv 1$, the loss reduces to the standard denoising objective.
  \item When the slate utility $U(\mathbf{x}_0|u,\mathbf{C})$ is not slot-decomposable, setting $r_j \equiv U(\mathbf{x}_0|u,\mathbf{C})/L$ recovers a utility-rewarded policy gradient.
  \item When $U$ is linear, i.e., $U = \sum_j r_j$, Eq.\eqref{eq:loss} simultaneously optimizes the average slate utility ($\frac{U}{\alpha L}\log p_\theta$) and an item-wise advantage term ($\frac{A_j}{\alpha}\log p_\theta$, where $A_j = r_j - U/L$), ensuring better alignment of item-level preferences with the context-aware model $p_\theta$.
\end{itemize}

\subsection{Inference}
\label{sec:method:variants}

At inference time, \name~starts from a fully masked slate and iteratively commits items to slots via the shared denoiser, as illustrated in Figure~\ref{fig:overview}(b). The inference behavior is primarily determined by the \emph{commit policy}: how many $\langle$slot, candidate$\rangle$ pairs to commit per step and whether the slot order is constrained across steps.

\textbf{Default \name~policy.}
At step $k$, we obtain the score matrix $\mathbf{Z}^{(k)}$ from Eq.\eqref{eq:latent-pool-score} and select the highest-confidence slot-candidate pair: 
\begin{equation}
(j^\star_k,\, i^\star_k) \;=\; \operatorname*{arg\,max}_{(j,\,i)}\; 
\mathbf{Z}^{(k)}_{j,\,i},
\label{eq:argmax}
\end{equation}
writing $\mathbf{c}_{i^\star_k}$ into slot $j^\star_k$, for $k = 1, \dots, L$. This confidence-ordered strategy makes easy decisions first and leaves uncertain slots for later rounds, when more slate context is available. After the $L$ denoising steps, an optional $T \geq 0$ refinement steps follow, where each step remasks the lowest-confidence item and reapplies the same denoiser (see \textbf{Appendix~\ref{app:refinement}}).

\textbf{Relation to existing paradigms.}
If the commit policy fills one pre-specified slot per step in 
left-to-right order, the framework recovers AR-style serial decoding. If it fills all slots in a single denoising pass, it recovers NAR-style parallel decoding. In all cases, \name~uses the same bidirectional denoiser and candidate-constrained interface (only the commit policy changes). More details are given in \textbf{Appendix~\ref{app:relations}}.

\section{Experiments}
\label{sec:exp}

\textbf{Datasets and Evaluation Metrics.} 
We conduct experiments on two public benchmarks and one industrial dataset: Amazon Books~\cite{mcauley2015image}, MovieLens-1M~\cite{harper2015movielens}, and Industry. 
Industry is a proprietary log collected from a real-world short-video platform, containing anonymized user interaction sequences. Dataset statistics are reported in \textbf{Appendix~\ref{app:dataset_statistics}}. 
For each dataset, 80\% of the data is used for training and 20\% for testing. Each data sample consists of a history of $|H|=50$ most recent interactions, a candidate pool of $|C|=50$ items, and the reranker output lists of length $L=6$ along with binary labels representing user feedback.
We report four list-wise metrics computed on the top-$L$ positions of the predicted list: Precision@$L$, NDCG@$L$, MAP@$L$, and F1@$L$. All metrics are higher-is-better and are averaged over requests in the test set.

\textbf{Baselines.} We consider nine reranking baselines spanning two families that dominate the recent literature. The \emph{Generator-only (G-only)} family trains a single generator to produce the output list end-to-end: DNN~\cite{covington2016deep}, DLCM~\cite{ai2018learning}, PRM~\cite{pei2019personalized}, GoalRank~\cite{zhang2026goalrank}, and GloRank~\cite{jia2026glorank}. The \emph{Generator-Evaluator (G-E)} family decouples list generation from list scoring so that a separately trained evaluator can rescore candidate lists produced by the generator: PIER~\cite{shi2023pier}, NAR4Rec~\cite{ren2024non}, G-$n$~\cite{yang2025comprehensive} with $n \in \{3, 20, 100\}$, and DCDR~\cite{lin2024discrete}. 
Full descriptions for baselines are provided in \textbf{Appendix~\ref{app:baselines}}.

\textbf{Implementation Details.} For Semantic ID construction, item text descriptions are encoded by Qwen3-4B into dense embeddings, reduced by PCA to $\texttt{128}$ dimensions, and then quantized by a multi-head VQ-VAE~\cite{shi2025llada} with $\texttt{M=4}$ disjoint codebooks of $\texttt{256}$ entries each. The backbone is a stack of $\texttt{4}$ bidirectional Transformer blocks with a hidden width of $\texttt{d=256}$, $\texttt{4}$ attention heads, and a dropout rate of $\texttt{0.1}$. We optimize the objective in Equation~\ref{eq:loss} with AdamW at a peak learning rate of $3 \times 10^{-3}$, weight decay of $\texttt{0.05}$, batch size of $\texttt{64}$, and \texttt{bf16} mixed precision. At each training step, for each example, we draw a mask count $n \sim \operatorname{Uniform}\{1, \dots, L\}$ and mask $n$ slate slots uniformly at random. We set the label-aware weight $r_j = 1$ for items with positive feedback (click or long watch, depending on the dataset) and $r_j = 0$ for the remaining slots. In inference, we run the default \name~confidence ordered denoising policy for $L$ steps, committing one item per step. We conduct all experiments on 2 NVIDIA RTX PRO 6000 Blackwell GPUs.

\subsection{Main Results}
\label{sec:exp:main}

\definecolor{impcolor}{RGB}{226, 240, 217}

\begin{table}[t]
\centering
\caption{Overall performance comparison on Amazon Books, MovieLens-1M, and Industry. All metrics are reported in percentage with $L=6$. The \textbf{best} and \underline{second-best} results are highlighted. The last two rows mark our method and its relative improvement over the strongest baseline.}
\label{tab:main-results}
\setlength{\tabcolsep}{3pt}
\renewcommand{\arraystretch}{1.05}
\resizebox{\textwidth}{!}{%
\begin{tabular}{lccccccccccccc}
\toprule
 & & \multicolumn{4}{c}{Amazon Books} & \multicolumn{4}{c}{MovieLens-1M} & \multicolumn{4}{c}{Industry} \\
\cmidrule(lr){3-6} \cmidrule(lr){7-10} \cmidrule(lr){11-14}
Family & Method & P@$L$ & NDCG@$L$ & MAP@$L$ & F1@$L$ & P@$L$ & NDCG@$L$ & MAP@$L$ & F1@$L$ & P@$L$ & NDCG@$L$ & MAP@$L$ & F1@$L$ \\
\midrule
\multirow{5}{*}{\shortstack[l]{G-only}}
 & DNN & 60.28 & 69.61 & 58.58 & 62.45 & 56.86 & 70.30 & 59.28 & 62.16 & 32.80 & 51.26 & 41.02 & 38.97 \\
 & DLCM & 66.80 & 75.88 & 65.39 & 69.28 & 62.31 & 73.87 & 63.82 & 67.96 & 47.86 & 73.03 & 63.16 & 56.89 \\
 & PRM & 67.86 & 76.88 & 66.44 & 70.42 & 60.09 & 72.85 & 62.21 & 65.51 & 47.89 & 67.09 & 54.59 & 56.88 \\
 & GoalRank & 80.35 & 84.88 & 77.91 & 83.44 & 73.56 & 83.43 & 76.16 & 80.15 & 60.16 & 88.08 & 82.06 & 71.48 \\
 & GloRank & \underline{83.75} & \underline{90.08} & \underline{85.10} & \underline{86.97} & \underline{75.79} & \underline{87.56} & \underline{81.19} & \underline{82.57} & \underline{62.56} & \underline{90.15} & \underline{84.84} & \underline{74.32} \\
\midrule
\multirow{6}{*}{\shortstack[l]{G-E}}
 & PIER & 71.14 & 80.22 & 71.62 & 73.74 & 62.74 & 75.99 & 65.98 & 68.74 & 56.69 & 80.15 & 70.52 & 67.52 \\
 & NAR4Rec & 70.08 & 79.46 & 70.69 & 72.66 & 62.81 & 75.01 & 65.42 & 68.31 & 51.93 & 65.16 & 53.13 & 61.56 \\
 & G-3 & 68.76 & 76.36 & 65.82 & 71.33 & 55.51 & 67.39 & 55.52 & 55.51 & 48.53 & 67.38 & 54.89 & 57.67 \\
 & G-20 & 72.99 & 78.68 & 68.66 & 75.72 & 58.66 & 69.86 & 58.60 & 64.18 & 51.58 & 69.29 & 57.26 & 61.08 \\
 & G-100 & 77.21 & 82.15 & 73.78 & 80.09 & 60.64 & 70.97 & 59.93 & 66.29 & 53.21 & 70.67 & 58.80 & 63.06 \\
 & DCDR & 80.56 & 86.73 & 80.33 & 83.66 & 73.04 & 84.61 & 77.17 & 79.60 & 50.02 & 62.48 & 49.26 & 59.44 \\
\midrule
\rowcolor{results}
 & \textbf{\name (Ours)} & \textbf{88.30} & \textbf{94.00} & \textbf{90.58} & \textbf{91.70} & \textbf{76.88} & \textbf{88.65} & \textbf{82.64} & \textbf{83.59} & \textbf{63.54} & \textbf{91.34} & \textbf{86.53} & \textbf{75.49} \\
\rowcolor{results}
 & \textit{Rel. Improv.} & +5.43\% & +4.35\% & +6.44\% & +5.44\% & +1.44\% & +1.24\% & +1.79\% & +1.24\% & +1.57\% & +1.32\% & +1.99\% & +1.57\% \\
\bottomrule
\end{tabular}%
}
\end{table}

Table~\ref{tab:main-results} reports the overall comparison between \name~and the reranking baselines on three datasets. We summarize three observations below:

\begin{itemize}[leftmargin=1.2em, itemsep=2pt, topsep=2pt]
    \item \textbf{\name~achieves the best overall performance.} It ranks first on all four metrics for three datasets. The gains are especially clear on Amazon Books, where \name~improves over the strongest baseline GloRank by 5.43\% in Precision, 4.35\% in NDCG, 6.44\% in MAP, and 5.44\% in F1. On MovieLens-1M and Industry, \name~also remains consistently above the best baseline method on every metric. This demonstrates the superiority of \name~and verifies the effectiveness of the unified framework.

    \item \textbf{Both AR and NAR reranking paradigms have intrinsic bottlenecks.} 
    Existing AR rerankers such as GoalRank and GloRank can model inter-item dependencies within the exposure slate, achieving superior performance within baselines.
    Yet, they are constrained by unidirectional causal attention and suffer from error propagation during decoding, resulting in suboptimal performance compared to \name.
    Existing NAR rerankers (DNN, DLCM, PRM, NAR4Rec) ignore the exposure dependency modeling and thus generally perform worse than AR rerankers in practice.
    DCDR uses diffusion-based refinement and achieves inter-item relation modeling, but its performance is also suboptimal compared to \name~due to its inherent dependency on the initial slate quality.

    \item \textbf{Larger evaluator search alone does not remove the gap to \name.} 
    Empirically, NAR rerankers perform better in the G-E paradigm (PIER and NAR4Rec) than in the G-only paradigm (DNN, DLCM, and PRM).
    This shows that the limitation caused by ignoring the exposure dependency modeling could be partially alleviated by exploring multiple slates as candidates, then using a slate-wise evaluator to select the best one.
    Within the G-$n$ variants, enlarging the slate candidate pool from G-3 to G-20 and G-100 steadily improves NDCG on Amazon Books from 76.36 to 78.68 and 82.15. However, even the strongest variant G-100 still remains well below \name~at 94.00. Similar gaps also appear on MovieLens-1M and Industry. 
    This result suggests that better evaluator side search cannot efficiently compensate for weaker slate generation.

\end{itemize}

It is worth noting that we also provide simulator-based experimental results~\cite{zhao2023kuaisim} in \textbf{Appendix~\ref{sec:exp:simu}} from a complementary perspective.

\subsection{Ablation Study}
\label{sec:exp:ablation}

We conduct an ablation study to examine the contribution of each component in \name. The variants fall into two groups: 

\begin{itemize}[leftmargin=1.2em, itemsep=2pt, topsep=2pt]
  \item \textbf{Input Components.}
  (1) w/o SID, replacing SID with item embeddings; (2) w/o Hist.\ Padding, removing history padding.

  \item \textbf{Model Architecture.}
  (1) w/o SFL, removing the Semantic Fusion Layer and feeding raw token embeddings into the Transformer; (2) w/o Iter.\ Decoding, collapsing the iterative denoising process into a single parallel step; and (3) w/o LPS, replacing Latent Pool Selection with direct SID token generation over the entire vocabulary.
\end{itemize}

\begin{table}[t]
\centering
\caption{Ablation study on Amazon Books and MovieLens-1M. All metrics are reported in percentage with $L=6$. The full model is shown at the top for reference.}
\label{tab:ablation}
\setlength{\tabcolsep}{4pt}
\renewcommand{\arraystretch}{1.05}
\resizebox{0.8\textwidth}{!}{%
\begin{tabular}{lcccccccc}
\toprule
 & \multicolumn{4}{c}{Amazon Books} & \multicolumn{4}{c}{MovieLens-1M} \\
\cmidrule(lr){2-5} \cmidrule(lr){6-9}
Method & P@$L$ & NDCG@$L$ & MAP@$L$ & F1@$L$ & P@$L$ & NDCG@$L$ & MAP@$L$ & F1@$L$ \\
\midrule
\name~(full) & \textbf{88.30} & \textbf{94.00} & \textbf{90.58} & \textbf{91.70} & \textbf{76.68} & \textbf{88.65} & \textbf{82.64} & \textbf{83.59} \\
\midrule
\multicolumn{9}{l}{\textit{Input Components}} \\
\midrule
\quad w/o SID & \underline{88.27} & 93.84 & 90.38 & \underline{91.65} & 76.33 & 88.17 & 82.08 & 83.19 \\
\quad w/o Hist.\ Padding & 88.25 & 93.78 & 90.32 & 91.64 & \underline{76.35} & \underline{88.41} & \underline{82.38} & \underline{83.23} \\
\midrule
\multicolumn{9}{l}{\textit{Model Architecture}} \\
\midrule
\quad w/o SFL & 81.13 & 88.31 & 82.24 & 84.25 & 75.21 & 86.69 & 79.94 & 81.96 \\
\quad w/o Iter.\ Decoding & 88.20 & \underline{93.89} & \underline{90.42} & 91.59 & 76.22 & 88.29 & 82.16 & 83.08 \\
\quad w/o LPS & 87.24 & 92.93 & 89.07 & 90.61 & 75.12 & 87.28 & 80.83 & 81.89 \\
\bottomrule
\end{tabular}%
}
\end{table}

Table~\ref{tab:ablation} reports the results, and we can draw the following conclusions:

\textbf{Input Components.} (1) \textit{Semantic ID provides marginal but consistent gains.} Replacing SID with raw item IDs (w/o SID) leads to consistent drops across all metrics. Semantic IDs encode structural item features through quantization, providing richer signals than opaque item indices. (2) \textit{History padding stabilizes the input context.} Removing history padding (w/o Hist.\ Padding) leads to small but consistent drops, particularly in MAP. History padding ensures that the Transformer attends over a fixed-length user interaction context, reducing variance introduced by requests with shorter histories.

\textbf{Model Architecture.} 
(1) \textit{The Semantic Fusion Layer (SFL) is essential for aligning item representations with the denoising process.} Removing SFL (w/o SFL) causes the largest drops across all settings, with Precision on Amazon Books falling from 88.30 to 81.13 and MAP dropping from 90.58 to 82.24. This breaks the alignment between denoising positions and items, making it substantially harder for the model to perform item-level scoring during the diffusion process. As a consequence, we suggest that practitioners should favor item-level modeling rather than token SID-level modeling when using diffusion in the reranking task.
(2) \textit{Iterative decoding provides consistent improvements over single-step generation.} 
A single parallel generation step (w/o Iter. Decoding) corresponds to the NAR extreme of \name, which performs slightly worse than the iterative decoding strategy due to the downgraded modeling of exposure dependency in the slate.
Fortunately, the performance drop is not severe; for systems operating under strict latency constraints, \name~w/o Iterative Decoding remains a highly competitive choice, as it avoids the computational overhead of the iterative generation process.
(3) \textit{Latent Pool Selection (LPS) is important for candidate-constrained generation.} Replacing LPS with direct SID token generation (w/o LPS) leads to notable drops, for example from 88.30 to 87.24 in Precision and from 90.58 to 89.07 in MAP on Amazon Books. Without LPS, the model must produce valid SIDs through a large shared vocabulary rather than selecting directly from the local candidate pool, which introduces errors from out-of-pool generation and weakens item comparison within a request.

\subsection{Denoising Parameter Analysis}
\label{sec:exp:denoising}

\begin{figure}
    \centering
    \includegraphics[width=\linewidth]{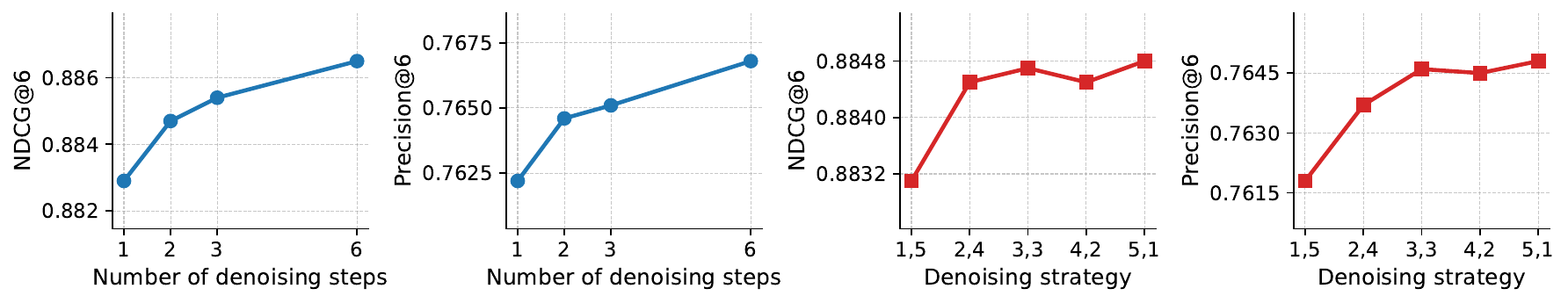}
    \caption{Effect of Denoising Parameters.}
    \label{fig:denoising_param_main}
\end{figure}

As shown in Figure~\ref{fig:denoising_param_main}, we investigate how the number of denoising steps and the denoising strategy affect the performance. Full results across all four metrics are provided in \textbf{Appendix~\ref{appendix:denoising_param}}.

\textbf{Number of denoising steps.} We vary the number of denoising steps among ${1, 2, 3, 6}$, where a step count of 1 means the model generates the entire list in a single inference pass, and a step count of 6 means each step generates exactly one item. Both NDCG and Precision improve consistently as the step count increases. This behavior resembles test-time scaling in large language models: each additional inference step allows the model to condition on items already placed in prior steps, providing stronger contextual signals when resolving positions with lower prediction confidence.

\textbf{Denoising strategy.} We fix the number of denoising steps at 2 and vary the allocation of items across the two steps. The label $a, b$ denotes generating $a$ items in the first step and $b$ items in the second step. Performance generally improves as more items are assigned to the first step. This is consistent with the confidence-based ordering underlying our denoising process: the first step resolves the positions with the highest confidence, where the model predictions are already reliable and additional inference budget yields limited gain. The remaining positions, denoised in the second step, have lower confidence and benefit more from the contextual information established in the first step. 

\subsection{Scaling Analysis}

\setlength{\intextsep}{5pt}
\setlength{\columnsep}{10pt}
\begin{wrapfigure}{r}{0.5\textwidth}
    \centering
    \includegraphics[width=\linewidth]{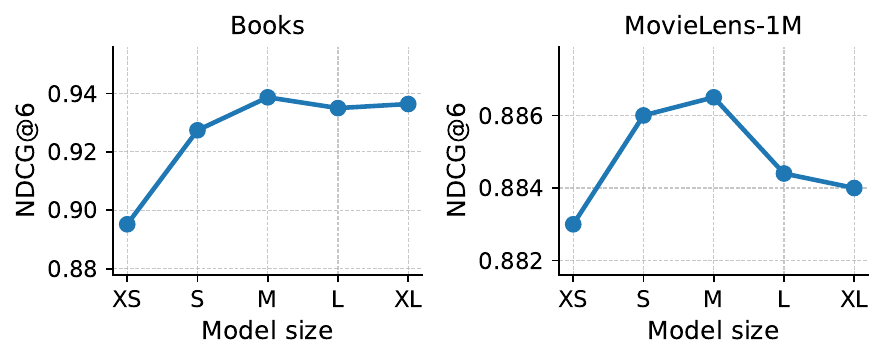}
    \caption{Scaling analysis with five configurations.}
    \label{fig:scaling}
\end{wrapfigure}
We examine how model size affects recommendation performance to identify the effective operating range of \name.
We evaluate five model configurations on both Books and MovieLens-1M: XS (0.29M), S (0.88M), M (5.08M, default), L (11.01M), and XL (27.59M) parameters. The complete results are provided in \textbf{Appendix~\ref{app:scaling}}.
As shown in Figure~\ref{fig:scaling}, performance improves consistently from XS to M on both datasets, then declines as model size increases further. This rise-then-fall pattern suggests that the training data volume imposes a practical capacity ceiling: models beyond the M configuration have sufficient parameters to overfit the available training signal, which leads to degraded generalization. The 5.08M default configuration achieves the best trade-off between model expressiveness and data efficiency.

\subsection{Online Experiments}
\label{sec:exp:online}

To evaluate \name~in a real-world industrial environment, we conduct an online A/B test on a short-video platform with billion-scale daily user interactions. The production system follows a multi-stage recommendation pipeline, where retrieval and ranking first produce a candidate pool, and the reranking stage selects the final exposed slate. In this deployment, the reranking stage follows the same 50-to-6 setting as our offline evaluation. The resulting permutation space contains nearly $1.5 \times 10^{10}$ possible slates, making exhaustive enumeration impractical.
  \setlength{\intextsep}{5pt}
  \setlength{\columnsep}{10pt}
  \begin{wraptable}{r}{0.5\textwidth}
  \centering
  \caption{Online A/B test results.}
  \label{tab:online_result}
  \small
  \setlength{\tabcolsep}{3pt}
  \renewcommand{\arraystretch}{1.08}
  \begin{tabular}{llc}
  \toprule
  Category & Metric & Rel. Improv. \\
  \midrule
  \multirow{3}{*}{Platform}
   & Ovarall app-time & +0.146\%*  \\
   & User avg. app-time & +0.159\%* \\
   & Ovarall Realshow & +0.273\%* \\
   & Long term DAU & +0.085\% \\
  \midrule
  \multirow{6}{*}{Interaction}
   & Like-rate & +0.588\%* \\
   & Follow-rate & +0.842\%* \\
   & Comment-rate & +0.285\% \\
   & Share-rate & +1.016\%* \\
   & Hate UV & -2.038\% \\
   & Effective interests & +0.082\% \\
  \bottomrule
  \end{tabular}
  \end{wraptable}
\textbf{Experimental Setup.}
The control system is the deployed reranking system. Its reranker follows a G-$n$-style~\cite{yang2025comprehensive}, where multiple generators produce candidate slates and an evaluator selects the final slate. The generator module is mainly based on a GoalRank-style model~\cite{zhang2026goalrank}. In the treatment system, we keep the online pipeline unchanged and replace the generator module in the reranking stage with \name. This design isolates the effect of \name~under the same retrieval, ranking, and evaluator components.

\textbf{Traffic Split and Test Period.}
We randomly allocate 10\% of online traffic to the control group and another 10\% to the treatment group, ensuring comparable user distributions across both buckets. The experiment runs for 7 consecutive days. This duration allows us to measure both statistical significance and cross-day stability under normal traffic changes.

\textbf{Evaluation Metrics.} We report both platform-level metrics and user interaction metrics. The platform metrics include overall app-time, user average app-time, overall Realshow, and long-term DAU, reflecting the system's impact on broad usage patterns and platform health. The interaction metrics include like-rate, follow-rate, comment-rate, share-rate, hate UV, and effective interests, capturing fine-grained user feedback on recommended content. Together, these metrics assess whether the reranker improves user engagement while maintaining stable long-term platform behavior.

\textbf{Main Results.}
Table~\ref{tab:online_result} reports the relative improvements of \name~over the control system. Results marked with an asterisk (*) are statistically significant (p<0.05). The results show that replacing the GoalRank-style generator with \name~improves the online reranking system in practice. This confirms that the proposed \name~framework can be deployed in a production multi-stage recommender system and can provide gains in real-world environment.

\section{Related Work}

\subsection{Reranking in Recommender Systems}\label{sec: related_work_reranking}

Reranking is the final stage of a modern multi-stage recommender pipeline, and it determines the list that is ultimately shown to the user~\cite{liu2022rerankingsurvey}. Existing neural rerankers can be broadly grouped into two paradigms: Generator-only and Generator-Evaluator. Generator-only methods use a single model to produce the slate directly~\cite{ai2018learning,bello2018seq2slate,pei2019personalized,zhang2026goalrank}. DLCM~\cite{ai2018learning} refines pointwise scores with a recurrent model that captures full list context; Seq2Slate~\cite{bello2018seq2slate} formulates slate construction as autoregressive decoding with a pointer network; PRM~\cite{pei2019personalized} replaces recurrent encoding with self-attention so that each candidate can interact with all other candidates in one pass; and GoalRank~\cite{zhang2026goalrank} revisits this paradigm with a large generator trained by group relative optimization. Generator-Evaluator methods~\cite{shi2023pier,ren2024non,yang2025comprehensive} separate list proposal from list scoring. PIER~\cite{shi2023pier} combines a SimHash-based permutation generator with a context-aware evaluator, NAR4Rec~\cite{ren2024non} generates slot assignments in parallel with unlikelihood training and contrastive decoding, and MG-E~\cite{yang2025comprehensive} uses multiple complementary generators to enlarge the candidate list space. Despite these design differences, both paradigms still rely on a fixed decoding order, which is either autoregressive and vulnerable to error propagation or fully parallel and therefore limited in modeling dependencies among target positions. \name~instead formulates reranking as masked discrete diffusion over the slate and commits slot to candidate assignments in order of their confidence, which unifies list generation and refinement within a unified framework and yields bidirectional slate attention together with an adaptive any order decoding schedule. The closest prior work to \name~is DCDR~\cite{lin2024discrete}, which is equivalent to the refinement stage of \name~applied to an external given initial list. As a result, its output is anchored to that initial list and its quality is tightly bounded by the quality of the starting point. 

\subsection{Generative Methods in Recommender Systems}\label{sec: related_work_generative}

Generative recommendation reformulates recommendation as a sequence generation problem over learned item identifiers, where the model predicts the next item token by token or through iterative denoising instead of assigning a pointwise matching score~\cite{li2025survey}. Progress in this direction can be organized along three axes. (1) Tokenization. Recent work replaces atomic item identifiers with content-informed representations. P5~\cite{geng2022recommendation} converts user item interactions, item metadata, and related signals into natural language sequences within a unified text-to-text framework, while TIGER~\cite{rajput2023recommender} learns hierarchical semantic identifiers from pretrained item embeddings through residual vector quantization. (2) Architecture. Encoder-decoder models such as M6-Rec~\cite{cui2022m6} use a unified pretrained language model to support retrieval, ranking, and explanation, whereas decoder-only models emphasize industrial scaling, with HSTU~\cite{zhai2024actions} proposing a high-throughput sequential transduction architecture for long and dynamic user histories. More recently, LLaDA-Rec~\cite{shi2025llada} replaces left-to-right decoding with a masked discrete diffusion backbone~\cite{nie2025llada} that generates the semantic identifier of the next item. (3) Optimization. Beyond next token prediction, TallRec~\cite{bao2023tallrec} shows that instruction tuning with small amounts of recommendation data can improve few-shot recommendation and cross-domain transfer. In contrast to these methods, which generate the next item from a global item vocabulary conditioned on user history, \name~targets the reranking setting, where the output must be selected from a small request-specific candidate pool.

\section{Conclusion}

We propose UniRank, a unified reranking framework that formulates slate generation as confidence-ordered denoising, bridging the tradeoff between dependency modeling and generation stability in existing AR and NAR rerankers.
We further introduce the Task Grounded Diffusion Interface, which aligns diffusion-based generation with reranking task.
Experiments on three datasets show that \name~consistently outperforms strong reranking baselines.

\newpage

\bibliography{reference}{}
\bibliographystyle{unsrt}

\clearpage
\appendix
\begin{center}
    \LARGE \textbf{Technical Appendix}
    \vspace{1em}
\end{center}
\addcontentsline{toc}{part}{Appendix}

\begingroup
  \etocsetnexttocdepth{subsection}
  \etocsettocstyle{\section*{Table of Contents}}{}
  \localtableofcontents
\endgroup
\clearpage

\section{Optional Refinement Stage}
\label{app:refinement}

\paragraph{Refinement procedure.}
We describe the optional refinement stage used after the insertion process in \name. After inference finishes, the model has produced a
complete slate $S=[s_1,\dots,s_L]$, where each $s_j$ is one candidate item from $C$. Refinement reuses the same denoising model and the same
Latent Pool Selection module, without additional training. At each refinement step, we reconsider one filled slot by a leave-one-out denoising
query. For slot $j$, we replace its current item representation with $e_{\text{mask}}$ and keep all other filled slots unchanged, which gives
a corrupted slate $S_{\setminus j}$. We then run the denoiser on $[\,H\,\mid\, e_{\text{sep}}\,\mid\, C\,\mid\, e_{\text{sep}}\,\mid\,
S_{\setminus j}\,]$ and obtain a candidate score vector $Z_{j,:}$ for the masked slot, following Equation~\ref{eq:latent-pool-score}. The
valid replacement set contains unused candidates in $C$ and the current item $s_j$, so the refined slate still contains distinct items, and the
model can also keep the original choice.

We use two criteria to decide which slot to reconsider. Let
\begin{equation}
p_{j,i}=\operatorname{softmax}(Z_{j,:})_i
\end{equation}
be the probability assigned to candidate $c_i$ when slot $j$ is masked, and let $i_j$ be the candidate index of the current item $s_j$. The
\textit{prob} criterion treats $p_{j,i_j}$ as the confidence of the current item and selects the slot with the lowest confidence. The
\textit{delta} criterion computes the replacement margin
\begin{equation}
\delta_j = \max_{i \in \mathcal{A}_j} p_{j,i} - p_{j,i_j},
\end{equation}
where $\mathcal{A}_j$ is the valid replacement set for slot $j$, and selects the slot with the largest margin. After selecting a slot, refinement writes the highest-scoring valid candidate into that slot. This procedure is repeated for a fixed number of refinement steps.

\begin{table}[t]
\centering
\caption{Effect of optional refinement on ML-1M. The insertion-only row is the default UniRank inference policy. Refinement reuses the same
denoising model at inference time. Higher values are better for all metrics.}
\label{tab:refinement}
\begin{tabular}{llcccc}
\toprule
Criterion & Steps & Precision@6 & NDCG@6 & MAP@6 & F1@6 \\
\midrule
Insertion only & -- & 0.7668 & 0.8865 & 0.8264 & 0.8359 \\
\midrule
\textit{prob} & 1 & 0.7668 & 0.8865 & 0.8267 & 0.8359 \\
\textit{prob} & 2 & \textbf{0.7671} & 0.8868 & 0.8267 & \textbf{0.8363} \\
\textit{prob} & 3 & \textbf{0.7671} & 0.8865 & 0.8267 & 0.8363 \\
\textit{prob} & 4 & 0.7669 & 0.8867 & 0.8266 & 0.8360 \\
\textit{prob} & 5 & 0.7665 & 0.8866 & 0.8266 & 0.8356 \\
\textit{prob} & 6 & 0.7667 & 0.8860 & 0.8261 & 0.8358 \\
\midrule
\textit{delta} & 1 & 0.7666 & 0.8864 & 0.8265 & 0.8356 \\
\textit{delta} & 2 & 0.7666 & 0.8863 & 0.8261 & 0.8357 \\
\textit{delta} & 3 & 0.7666 & 0.8858 & 0.8257 & 0.8356 \\
\textit{delta} & 4 & 0.7667 & 0.8862 & 0.8260 & 0.8358 \\
\textit{delta} & 5 & 0.7670 & 0.8868 & 0.8268 & 0.8362 \\
\textit{delta} & 6 & 0.7667 & \textbf{0.8869} & \textbf{0.8272} & 0.8358 \\
\bottomrule
\end{tabular}
\end{table}

\paragraph{Analysis.}
Table~\ref{tab:refinement} reports the refinement results on ML-1M. Refinement gives small changes over the insertion-only result. With the
\textit{prob} criterion, two refinement steps give the best Precision@6 and F1@6, improving Precision@6 from 0.7668 to 0.7671 and F1@6
from 0.8359 to 0.8363. With the \textit{delta} criterion, six refinement steps give the best NDCG@6 and MAP@6, improving NDCG@6 from
0.8865 to 0.8869 and MAP@6 from 0.8264 to 0.8272.

The results indicate that the insertion process already produces stable slates, while refinement can make inference-time corrections
by rechecking filled slots under bidirectional slate context. The gains are not uniform across metrics or step counts. More refinement steps
do not always improve the result, which suggests that repeatedly replacing filled items may also disturb correct earlier choices. We therefore
treat refinement as an optional inference-time extension of UniRank, rather than a required part of the main method.

\section{Relation to Existing Paradigms}\label{app:relations}

\textbf{Explaining existing works under \name~terminologies:} We provide detailed model settings of existing methods and \name~in Table \ref{tab:unirank_variation}.

\begin{table}[t]
  \centering
  \resizebox{\linewidth}{!}{
  \begin{tabular}{c|c|c|c|c|c}
      \toprule
      \multicolumn{2}{c|}{Model} & \#step &
      \begin{tabular}{c}
          \#token \\ per step
      \end{tabular}
      & Commit policy & Training objective \\
      \midrule
      \multirow{3}{*}{AR}
      & Seq2Slate~\cite{bello2018seq2slate} & $L$ & 1 & $P(s_k \mid s_{<k}, u, C)$ & Step-wise sequence CE \\
      & GoalRank~\cite{zhang2026goalrank} & $L$ & 1& $P(s_k \mid s_{<k}, u, C)$ & Group-relative CE/KL \\
      & GloRank~\cite{jia2026glorank} & $L\times M$ & 1 & $P(\text{sid}_k \mid \text{sid}_{<k}, u, C)$ & NTP CE + GRPO \\
      \midrule
      \multirow{3}{*}{NAR}
      & NAR4Rec~\cite{ren2024non} & 1 & $L$ & Parallel generation & Unlikelihood + contrastive losses \\
      & PRM~\cite{pei2019personalized} & 1 & $L$ & top-$L$ ranking & Click CE/NLL \\
      & DNR~\cite{mao2026denoising} & 1 & $L$ & top-$L$ ranking & $\max P(Y \mid \tilde{Y}, u, C, S)$ \\
      \midrule
      \multicolumn{2}{c|}{UniRank(Default)} & $L+T$ & 1 & \multirow{2}{*}{$\operatorname*{arg\,max}_{(j,i)}\, Z_{j,i}$} & \multirow{2}{*}{$\frac{r_j}
{\alpha}\text{CE}(P(i_j \mid x_\alpha, u, C), y_j)$} \\
      \multicolumn{2}{c|}{UniRank(NAR)} & 1 & $L$ & & \\
      \bottomrule
  \end{tabular}}
  \vspace{3pt}
  \caption{Model variations under the \name~framework.}
  \label{tab:unirank_variation}
\end{table}

\section{Dataset Statistics}
\label{app:dataset_statistics}

As noted in the experimental setup, the UniRank experiments use three datasets: MovieLens-1M, Amazon Books, and Industry Dataset.
Table~\ref{tab:dataset_statistics} reports the number of users, items, interactions, and lists for each dataset.

\begin{table}[t]
\centering
\caption{Dataset statistics.}
\label{tab:dataset_statistics}
\resizebox{0.65\linewidth}{!}{
\begin{tabular}{lcccc}
\toprule
Dataset            & Users   & Items  & Interactions & Lists    \\
\midrule
MovieLens-1M       & 6,020   & 3,043  & 995,154      & 161,646  \\
Amazon Books       & 35,732  & 38,121 & 1,960,674    & 311,386  \\
Industry           & 200,775 & 17,014 & 5,173,698    & 589,751  \\
\bottomrule
\end{tabular}}
\end{table}

\section{Baseline Descriptions}
\label{app:baselines}

In this section, we detail the descriptions of each baseline in our experiments.

\paragraph{Generator-only baselines.}
These methods directly produce the output list or assign final item scores with a single model.

\begin{itemize}[leftmargin=*]

    \item \textbf{DNN}~\cite{covington2016deep} is adapted from the deep ranking model in the YouTube recommendation system. It scores each candidate item independently with a neural network and ranks items by the predicted scores, without explicit list-level dependency modeling.
    
    \item \textbf{DLCM}~\cite{ai2018learning} is a listwise context model originally proposed for ranking refinement in information retrieval. It sequentially encodes the top-ranked items with a recurrent neural network and uses the learned local ranking context to adjust item scores.
    
    \item \textbf{PRM}~\cite{pei2019personalized} is a personalized reranking model based on a Transformer encoder. It models the mutual influence among items in the input list and incorporates personalized vectors to capture user-specific preferences before producing reranking scores.
    
    \item \textbf{GoalRank}~\cite{zhang2026goalrank} is a generator-only one-stage ranking framework for training a large ranker with group-relative optimization. It constructs a reference policy from reward-model scores within candidate-list groups and trains the generator to
    align with this group-relative policy.
    
    \item \textbf{GloRank}~\cite{jia2026glorank} reformulates reranking from selecting local candidate indices to generating global item
    identifiers represented by Semantic IDs. It uses constrained decoding over the candidate set to produce valid non-duplicated output lists.

\end{itemize}

\paragraph{Generator-Evaluator baselines.}
These methods decouple candidate-list generation from list scoring, where an evaluator selects the final list from generated candidates.

\begin{itemize}[leftmargin=*]
  \item \textbf{PIER}~\cite{shi2023pier} is a Generator-Evaluator reranking framework for e-commerce recommendation. It first selects promising candidate permutations according to user permutation-level interests and then evaluates these permutations with an omnidirectional context-aware prediction module.

  \item \textbf{NAR4Rec}~\cite{ren2024non} is a non-autoregressive generative model for reranking recommendation. It generates all target positions in parallel, which reduces the sequential decoding cost of autoregressive generators while retaining a Generator-Evaluator reranking structure.

  \item \textbf{G-$n$}~\cite{yang2025comprehensive} is a multi-generator reranking framework, where $n$ generators produce candidate lists that are then scored by an evaluator. We evaluate its variants with $n \in \{3, 20, 100\}$ to study the effect of expanding the generated candidate-list set.

  \item \textbf{DCDR}~\cite{lin2024discrete} applies discrete conditional diffusion to reranking. It defines a discrete forward process over item sequences and learns a conditional reverse process to refine reranked lists under expected user-feedback conditions.

\end{itemize}

\section{Simulator-based Evaluation}
\label{sec:exp:simu}

\begin{table}[t]
  \centering
  \caption{Simulator-based evaluation on the reranking task. We report the average reward, where higher is better. The \textbf{best} and \underline{second-best} results are highlighted.}
  \label{tab:simulator_eval}
  \small
  \setlength{\tabcolsep}{14pt}
  \begin{tabular}{lclc}
      \toprule
      Method & Reward (avg) $\uparrow$ & Method & Reward (avg) $\uparrow$ \\
      \midrule
      DNN & 1.6146 & MGE-3 & 1.7997 \\
      DLCM & 1.5920 & MGE-20 & 1.8878 \\
      PRM & 1.5796 & MGE-100 & 1.8914 \\
      NAR4Rec & 1.5723 & DCDR & 2.8200 \\
      PIER & 1.7837 & GloRank & \underline{2.9730} \\
      GoalRank & 1.8851 & \textbf{Ours} & \textbf{3.0551} \\
      \bottomrule
  \end{tabular}
\end{table}

To evaluate \name~from a complementary perspective, we assess the list-level utility of the reranked output using a user interaction simulator, which provides a direct measure of practical recommendation effectiveness. Following previous work~\cite{zhao2023kuaisim}, we train a list-wise simulator on MovieLens-1M and use it to score the output list of each method on the test set. \name~and each baseline method are pre-trained on the same training set. The average reward across all test instances is reported as the evaluation metric. As shown in Table~\ref{tab:simulator_eval}, \name~achieves the highest average reward of 3.055, outperforming the strongest baseline GloRank and DCDR. The remaining baselines, including GoalRank, G-100, and PIER, cluster in the range of 1.57 to 1.89. The result confirms that \name~generates lists with higher overall user utility, which is consistent with its discrete denoising process that reasons over the full list at each step.

\section{Additional Analysis on Denoising Parameters}
\label{appendix:denoising_param}

\begin{figure}[t]
  \centering
  \includegraphics[width=\linewidth]{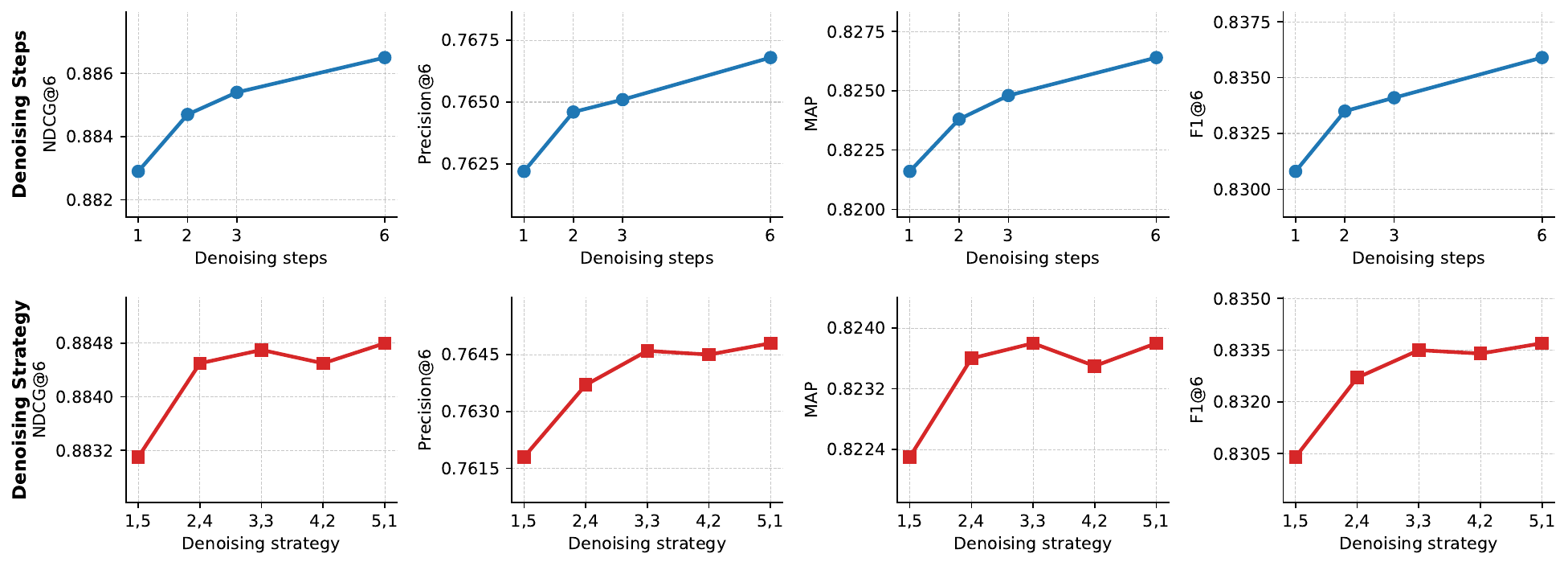}
  \caption{Effect of denoising parameters on all evaluation metrics. The upper row varies the number of denoising steps, and the lower row
varies the item allocation strategy under two denoising steps.}
  \label{fig:denoising_param}
\end{figure}

Figure~\ref{fig:denoising_param} reports the complete hyperparameter analysis for the denoising process. The upper row varies the number of denoising steps from 1 to 6. All four metrics improve as the number of steps increases, showing that iterative denoising helps refine the output list. The gain from 1 to 2 steps is clear, while later steps bring smaller additional improvements. The lower row studies the item allocation strategy under two denoising steps. The label $a,b$ means that $a$ items are generated in the first step and $b$ items are generated in the second step. The strategy $1,5$ gives the weakest results across all metrics. Moving more items to the first step generally improves performance, and $5,1$ achieves the best NDCG@6, Precision@6, and F1@6, while matching the best MAP. This trend supports the confidence-ordered denoising design, where high-confidence positions are fixed first, and the remaining lower-confidence positions are generated with the context established by the first step.

\section{Additional Scaling Results}
\label{app:scaling}

\begin{figure}[t]
  \centering
  \includegraphics[width=\linewidth]{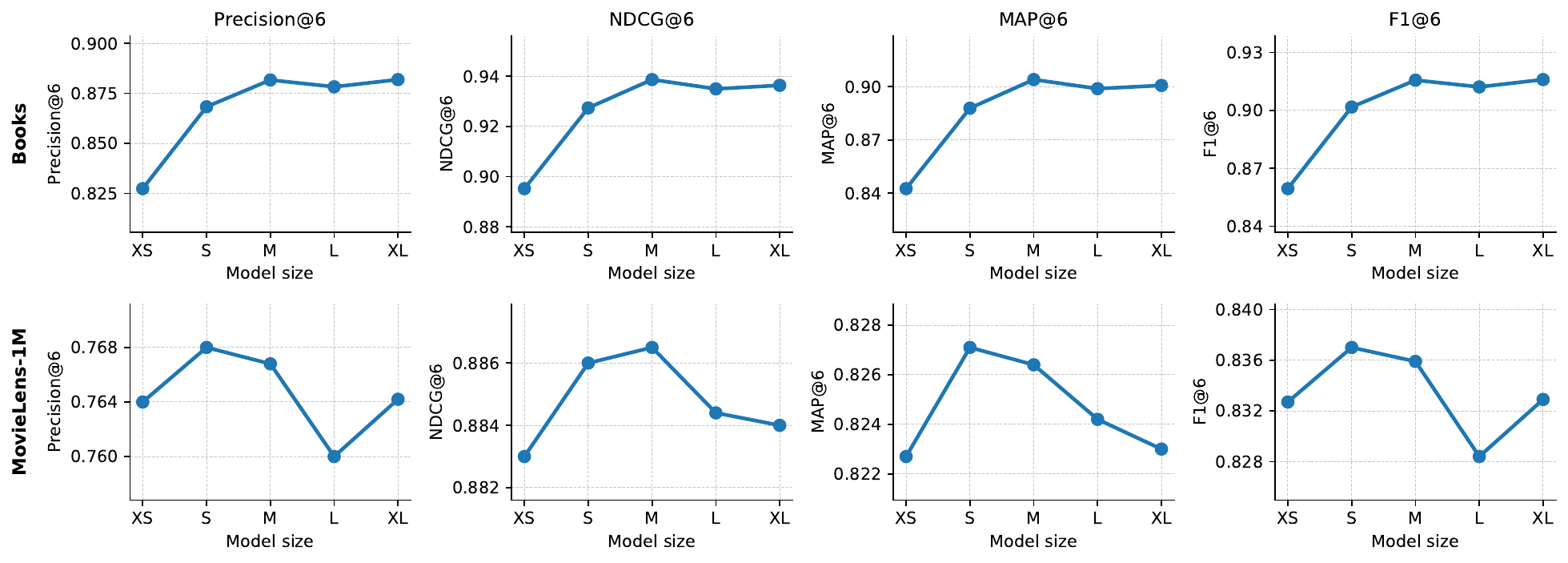}
  \caption{Complete scaling results on Books and MovieLens-1M.}
  \label{fig:scaling_appendix}
\end{figure}

Figure~\ref{fig:scaling_appendix} reports the complete scaling results on Books and MovieLens-1M across Precision@6, NDCG@6, MAP@6, and F1@6. We
evaluate five model configurations: XS with 0.29M parameters, S with 0.88M parameters, M with 5.08M parameters, L with 11.01M parameters, and XL with 27.59M parameters. On Books, performance improves substantially from XS to M across all metrics. Larger models do not bring consistent gains: L is lower than M on all four metrics, while XL only slightly improves Precision@6 and F1@6 but remains below M on NDCG@6 and MAP@6. On MovieLens-1M, the best results are concentrated around S and M, and both L and XL perform worse than these medium-size configurations. These results support the conclusion that increasing model capacity is useful up to a moderate scale, while larger models can overfit the available training signal and reduce generalization.

\section{Limitations}
\name~is designed for the reranking stage and therefore selects items from a request-specific candidate pool. This design matches the standard multi-stage recommendation pipeline, but it also means that the final slate quality is bounded by the upstream retrieval and ranking stages. A natural extension is to jointly optimize candidate generation and diffusion-based reranking, or to use \name~as feedback for improving the upstream candidate pool.

\end{document}